\documentclass[twoside,onecolumn]{revtex4}
\usepackage{bm}
\usepackage[utf8]{inputenc}
\usepackage{amssymb,theorem,xspace}
\usepackage{amsmath}
\usepackage{mathrsfs}
\usepackage{pgfplots}
\usepackage{graphicx}
\begin{document}
\title{The Landau Problem and non-Classicality}%
\author{Petronilo, G. X. A.}
\affiliation{ International Center of Physics, Universidade de Brasília, 70.910-900, Brasília, DF, Brazil }
\email{gustavo.petronilo@aluno.unb.br}
\author{Ulhoa, S. C.}
\affiliation{ International Center of Physics, Universidade de Brasília, 70.910-900, Brasília, DF, Brazil }
\affiliation{Canada Quantum Research Center,\\
204-3002 32 Ave Vernon, BC VIT 2L7, Canada}
\author{Araújo, K. V. S.}
\affiliation{ International Center of Physics, Universidade de Brasília, 70.910-900, Brasília, DF, Brazil }
\author{Paiva, R. A. S.}
\affiliation{ International Center of Physics, Universidade de Brasília, 70.910-900, Brasília, DF, Brazil }
\author{Amorim, R. G. G.}
\affiliation{ International Center of Physics, Universidade de Brasília, 70.910-900, Brasília, DF, Brazil }
\affiliation{Department of Physics, Faculdade Gama, University of Brasília, 71.910-900, Brasília, DF, Brazil}

\author{Santana, A. E.}
\affiliation{ International Center of Physics, Universidade de Brasília, 70.910-900, Brasília, DF, Brazil }

\begin{abstract}
Exploring the concept of the extended Galilei group G. Representations for field theories in a symplectic manifold have been derived in association with the method of the Wigner function. The representation is written in the light-cone of a de Sitter space-time in five dimensions. A Hilbert space is constructed, endowed with a symplectic structure, which is used as a representation space for the Lie algebra of G. This representation gives rise to the spin-zero Schrödinger (Klein-Gordon-like) equation for the wave functions in phase space, such that the dependent variables have the content of position and linear momentum. This is a particular example of a conformal theory, such that the wave functions are associated with the Wigner function through the Moyal product. We construct the Pauli-Schrödinger (Dirac-like) equation in phase space in its explicitly covariant form. In addition, we analyze the gauge symmetry for spin 1/2 particles in phase space and show how implement the minimal coupling in this case. We applied to the problem of an electron in an external field, and we recovered the non-relativistic Landau Levels. Finally, we study the parameter of negativity associated with the non-classicality of the system.
\end{abstract}
\maketitle

 \section{Introduction}

Galilean symmetry is the cornerstone of classical physics and today, even with the advent of discovery of others, more general symmetries, it is relevant in the research of various low-speed phenomena mainly associated with fluid theory.

Takahashi \emph{et. al.} introduced in 1988 the Galilean covariance with the purpose to construct covariant non-relativistic versions of field theories.  The Schrödinger equation was then written in a similar form as Klein-Gordon equation but in the light-cone of a $(4+1)$ de Sitter space~\cite{tak1,tak2,tak3}. With Galilean covariance it's possible to construct a tensor algebra~\cite{pinski1968} that gives rise to the study of a non-relativistic version of Dirac equation, closely related to the Pauli-Schrödinger equation. Such a  covariant version of non-relativistic equations provides a symmetry-consistent analysis of problems that solutions are not known in the usual formalism, such as the study of spin for low-velocity systems interacting with weak-gravitational fields~\cite{ulhoa2009galilean}. In similar cases, the Galilean covariance permits the breaking of spin degeneracy without the use of perturbation theory. One important but still little studied is the generalization of Galilean covariance to a  quantum phase space~\cite{petron23}. This is one of our goals here.

The first successful formalism for quantum mechanics in phase space was proposed by Wigner in 1932~\cite{wig0}, where the objective was to study kinetic theory in a quantum perspective~\cite{wig0,wig1,wig3}. Since then, the Wigner formalism has been applied in many areas, such as quantum computing, condensed matter, nuclear physics, and plasma physics~\cite{ap1}-\cite{ap14}. From a theoretical perspective, in the Wigner formalism an operator $A$ defined in Hilbert space $\mathcal{H}$ is associated to a function in a phase space, say $A \rightarrow a_W(q,p)$. The product of two operators, $AB$, both defined in $\mathcal{H}$, is associated to a star-product of their phase space correspondents, i.e.~\cite{wig2,wig3},
\begin{equation}\label{star}
(AB)_W\rightarrow a_W(q,p)\star b_W(q,p)=a_W(q,p) e^{\frac{i\hbar}{2}\left(\frac{\overleftarrow{\partial}}{\partial q}\frac{\overrightarrow{\partial}}{\partial p}-\frac{\overleftarrow{\partial}}{\partial p}\frac{\overrightarrow{\partial}}{\partial q}\right)} b_W(q,p).
\end{equation}
Such result can be seen as the action of a star-operator, $a_W(q,p)\star$, in the function $b_W(q,p)$. This fact is useful for define the star-operators $a_W(q,p)\star$,  which are used to build up representations of symmetry groups in a symplectic manifold. Using the star-operators, Oliveira et al studied a unitary representation of Galilei group and obtained the Schrödinger equation in phase space~\cite{ol1}.  In the same perspective, but studying relativistic systems, Amorim and co-workers. analyzed unitary representations of Poincaré group, writing the Klein-Gordon and Dirac equation in phase space~\cite{ron0}. In both relativistic or non-relativistic representations, the solutions of equations in phase space, $\psi(q,p)$ are related to Wigner function by the star-product, that is., $f_W(q,p)=\psi(q,p)\star\psi^{\dagger}(q,p)$. This provides a fully physical interpretations for the representation, with an interesting perspective: This formalism provides a way to address gauge theories and perturbative methods in phase space context,  which is not a simple task in the usual Wigner formalism~\cite{ron1,ron2,ron3}.

The Wigner function is the main object calculated in this article, for this we use the formalism of the phase space, whose projections on the axis of the momenta or coordinates, reproduce the results of the usual quantum mechanics.  Thus, the formulation of a given field in the phase space is more generic and allows a better understanding of the physical characteristics of that field.  In particular, the Wigner function in the context of the Pauli-Schr{\"o}dinger equation is calculated using Galilean covariance. With this, we analyze the negativity parameter of the system in order to establish its non-classicality~\cite{kenfack2004, quantumcomputing,delfosse2017,Rabbou2019}.

We develop and expand preliminary results in the study of the Galilean covariance in phase space~\cite{petron23}. Specifically, we obtain the symplectic version of the Pauli-Schrödinger equation. Using this equation, we studied the Landau problem in phase space, performing the calculation of the Wigner function and energy level for the system. In order to analyse non-classicality of the Landau states, we calculate the negativity parameter of the Wigner function.

The Galilean covariance was used in~\cite{tak2} to analyze the Pauli-Schr{\"o}dinger equation. We recommend the references there in. On the other hand, the formulation of this field in phase space is still not fully developed.  In this article we deal with this problem.

The presentation of this work is organized as follows. In Sec.~\ref{gc} the construction of the Galilean Covariance is presented.  Sec.~\ref{Galilei Group and Symplectic Quantum Mechanics}, a symplectic structure is constructed in the Galilean manifold. and using the commutation relations the Schrödinger equation in the light-cone of five dimensions in phase space is constructed, using a proposed solution the Schrödinger equation in phase space is restored to its non-covariant form in (3+1) dimensions. The explicitly covariant Pauli-Schrödinger equation is constructed in sec.~\ref{Spin 1/2 Symplectic Representaion}. In sec. IV we analyze the gauge symmetry for non-relativistic spin 1/2 particle in phase space. We study the Galilean spin 1/2 particle with an external field and solutions are proposed and discussed, after we calculated the negativity parameter and discuss the physical meaning. The sec.~\ref{Concluding Remarks} presents some concluding remarks.

\section{Galilean Covariance} \label{gc}

The Galilean transformations, that describes non-relativistic classical physics, are given by
\begin{eqnarray}
\overline{\mathbf{x}} &=&R\mathbf{x+v}t+\mathbf{a,}  \nonumber \\
\overline{t} &=&t+b.  \label{t gal2}
\end{eqnarray}%
where $R$ is the three-dimensional Euclidean rotations, $\textbf{v}$ stands for the relative velocity defining the Galilean boosts, $\textbf{a}$ is the spatial translations and $b$, is the time translations.
To obtain Galilean covariant transformations, consider the non-relativistic mass-shell relation
\begin{equation}
p_{\mu }p^{\mu }=\mathbf{p}^{2}-2mE=0,  \label{disp5}
\end{equation}
with $p^{\mu}=(\mathbf{p},p^{5}=E,p^{4}=m)$, $p_{\mu }=g_{\mu\nu}p^{\nu }$ and
\begin{equation}
g_{\mu\nu}= \left(
  \begin{array}{ccccc}
    1 & 0 & 0 & 0 & 0 \\
    0 & 1 & 0 & 0 & 0 \\
    0 & 0 & 1 & 0 & 0 \\
    0 & 0 & 0 & 0 & -1 \\
    0 & 0 & 0 & -1 & 0 \\
  \end{array}
\right) \,. \label{1}
\end{equation}
Here $ g_{\mu \nu} $ is the Galilean metric. There is a manifold equipped with a metric whose coordinates transform as follows
\begin{equation}
x^{\mu}\,'=\Lambda^{\mu}\,_{\nu}  x^{\nu}+a^\mu
\,,\label{3.2}
\end{equation}
where $a^\mu=({\bf a}, b,0)$ and $\Lambda^{\mu}\,_\nu $ is given by

\begin{equation}
\Lambda^{\mu}\,_\nu= 
\left(\begin{array}{ccccc}
   R^1_{\text{\;}1}&R^1_{\text{\;}2}&R^1_{\text{\;}3}&0&-v_1\\
   R^2_{\text{\;}1}&R^2_{\text{\;}2}&R^2_{\text{\;}3}&0&-v_2\\
   R^3_{\text{\;}1}&R^3_{\text{\;}2}&R^3_{\text{\;}3}&0&-v_3\\
   v_iR^i_{\text{\;}1}&v_iR^i_{\text{\;}2}&v_iR^i_{\text{\;}3}&1&\frac{\textbf{v}^2}{2}\\
      0&0&0&0&1
   \end{array}\right).
\end{equation}
The invariant line element is
\begin{equation}
dl^2=g_{\mu\nu}dx^{\mu}dx^{\nu}\,,\label{3.6}
\end{equation}
with $g_{\mu\nu}=\Lambda^{\alpha}\,_\mu \Lambda_{\alpha\nu}$. The fifth coordinate $ x^5 = s $ is canonically conjugate to the mass. Galilean physics lies on a null geodesic in 5-dimensional space. So $ dr^ 2-2dtds = 0 $, resulting in $ ds = \frac {{\bf v}} {2} \cdot{\bf dr} $. In this sense, the fifth coordinate is assigned to the other four. Therefore Galilean transformations in the equations~\eqref{t gal2} are considered to include 4-dimensional physical space in 5-dimensional space.
\section{The Extended Galilei Group in $\mathcal{H}(\Gamma)$ and Quantum Mechanics in Phase Space}\label{Galilei Group and Symplectic Quantum Mechanics}

To associate the Hilbert space, $\mathcal{H}$, with the phase space $\Gamma$, we consider the set of complex valued square-integrable functions, $\phi(q,p)$ in $\Gamma$, such that
\begin{equation}
\int dpdq\; \phi^{\ast}(q,p)\phi(q,p) <\infty.
\end{equation}
Then we can write $\phi(q,p)=\langle q,p|\phi\rangle$, with the aid of
\begin{equation}
\int dp dq |q,p\rangle\langle q,p| =1,
\end{equation}
where $\langle\phi|$ is the dual vector of $|\phi\rangle$. This symplectic Hilbert space is denoted by $H(\Gamma)$.

Now, we analyze the Galilei group considering $H(\Gamma)$ as a representation space. To do this, consider the unit transformations, $U\text{:}\mathcal{H}(\Gamma)\rightarrow\mathcal{H}(\Gamma)$, such that $\langle \psi_1|\psi_2\rangle$ is invariant.
Using the operator $\Lambda$, we define a mapping $e^{i\frac{\Lambda}{2}}=\star\text{:}\Gamma\times\Gamma\rightarrow\Gamma$ called as
\begin{eqnarray*}
f\star g=f(q,p)\,\,\exp\left[\frac{i}{2}\left(\frac{\overleftarrow{\partial}}{\partial q^\mu}\frac{\overrightarrow{\partial}}{\partial p_\mu}-\frac{\overleftarrow{\partial}}{\partial p^\mu}\frac{\overrightarrow{\partial}}{\partial q_\mu}\right)\right]g(q,p),\nonumber\\
\end{eqnarray*}
where $\hbar=1$.
To construct a representation of Galilei algebra in $\mathcal{H}$, we define the following operators,
\begin{subequations}
\begin{eqnarray}
\widehat{P}^\mu&=&p^\mu\star=p^\mu-\frac{i}{2}\partial_{q_\mu},\label{eq__p1}\\
\nonumber\\
\nonumber\\
\widehat{Q}^\mu&=&q^\mu\star=q^\mu+\frac{i}{2}\partial_{p_\mu}.
\end{eqnarray}
and\nonumber
\begin{eqnarray}
\widehat{M}_{\nu\sigma}&=&M_{\nu\sigma}\star=\widehat{Q}_\nu\widehat{P}_\sigma-\widehat{Q}_\sigma\widehat{P}_\nu.
\end{eqnarray}
\end{subequations}
Where $\widehat{M}_{\nu\sigma}$ are the generators of homogeneous transformations and $\widehat{P}_\mu$ of the non-homogeneous. From this set of unitary operators we obtain, after some simple calculations, the following set of commutations relations,
\begin{eqnarray*}
\left[\widehat{M}_{\mu\nu},\widehat{M}_{\rho\sigma}\right]&=&-i(g_{\nu\rho}\widehat{M}_{\mu\sigma}-g_{\mu\rho}\widehat{M}_{\nu\sigma}+g_{\mu\sigma}\widehat{M}_{\nu\rho}-g_{\mu\sigma}\widehat{M}_{\nu\rho}),\\
\nonumber\\
\left[\widehat{P}_\mu, \widehat{M}_{\rho\sigma}\right]&=&-i(g_{\mu\rho}\widehat{P}_\sigma-g_{\mu\sigma}\widehat{P}_\rho),\\
\nonumber\\
\left[\widehat{P}_\mu, \widehat{P}_{\sigma}\right]&=&0.
\end{eqnarray*}
These relations form a closed algebra which the Lie algebra of Galilei group in the case of $\mathcal{R}^3\times t$, is a subalgebra.
Considering $\widehat{J}_i=\frac{1}{2}\epsilon_{ijk}\widehat{M}_{jk}$ the generators of rotations and $\widehat{K}_i=\widehat{M}_{5i}$ of the pure Galilei transformations, $P_\mu$ the spatial and temporal translations. The commutation of $K_i$ and $P_i$ is naturally non-zero in this context, being $P_5$ related with mass. \vskip 2pt
The invariants of this algebra are
\begin{eqnarray}
I_1&=&\widehat{P}_\mu \widehat{P}^\mu\label{I-1}\\
I_2&=&\widehat{P}_5\label{I-2}
\end{eqnarray}
Using the Casimir invariants $ I_1 $ and $ I_2 $ and applying in $\Psi$, we have:
\begin{eqnarray*}\label{DKP-22}
   \begin{array}{ll}
     \widehat{P}_{\mu}\widehat{P}^{\mu}\Psi=k^2\Psi\label{eq:g1}\\
      \widehat{P}_{5}\Psi=-m\Psi
       \end{array}
\end{eqnarray*}
From this we obtain
\begin{equation*}
   \left(p^2-i\textbf{p}\cdot\boldsymbol{\nabla}-\frac{1}{4}\nabla^2-k^2\right)\Psi=\left(p_4-\frac{i}{2}\partial_t\right)\left(p_5 -\frac{i}{2}\partial_5\right)\Psi,
\end{equation*}
a solution for this equation is
\begin{eqnarray}
    \Psi=e^{-2i\left[(p_5+m)q_5+(p_4+E)t\right]}\Phi(q,p).
\end{eqnarray}
So, we have
\begin{equation*}
      \frac{1}{2m}\left(p^2-i\textbf{p}\cdot\boldsymbol{\nabla}-\frac{1}{4}\boldsymbol{\nabla}^2\right)\Phi=\Big(E+\frac{k^2}{2m}\Big)\Phi,
\end{equation*}
\vskip 8pt
which is the usual form of the Schrödinger equation in the phase space for the free particle with mass $m$~\cite{ol1}, with an additional kinetic energy of $\frac{k^2}{2m}$, that we can always set as the zero of energy.\vskip 2pt
This equation, and its complex conjugate, can be obtained by the Lagrangian density in phase space (we use $\partial^\mu=\partial/\partial q_\mu$)
\begin{eqnarray*}
    \mathcal{L}&=&\frac{1}{4}\partial^\mu\Psi(q,p)\partial\Psi^*(q,p)+\frac{i}{2}p^\mu[\Psi(q,p)\partial^\mu\Psi^*(q,p)-\Psi^*(q,p)\partial^\mu\Psi(q,p)]+\left[p^\mu p_\mu-k^2\right]\Psi.
\end{eqnarray*}
The association of this representation with the Wigner formalism is given by
\begin{eqnarray*}
f_w(q,p)=\Psi(q,p)\star\Psi^\dagger(q,p)
\end{eqnarray*}
where $f_w(q,p)$ is the Wigner function.

Which satisfies the 5-dimensional Galilean covariant Liouville-von Neumann equation in phase space, given by
\begin{eqnarray}
p_\mu\partial_{q_\mu}f_w(q,p)=0.\label{W-L}
\end{eqnarray}
The Pauli-Schrödinger equation in Galilean covariant formalism has the form of the Dirac equation,
\begin{equation}
    \left(\gamma^\mu\widehat{P}_\mu-k\right)\Psi(p,q)=0\label{eq:DPS}
\end{equation}
or
\begin{equation}
     \gamma^\mu\left(p_\mu-\frac{i}{2}\partial_\mu\right)\Psi(p,q)=k\Psi(p,q)
\end{equation}
Eq.~\eqref{eq:DPS} can  be derive from the Lagrangian density for spin 1/2 particles in phase space, which is given by
\begin{equation*}
\mathcal{L}=-\frac{i}{4}\Big((\partial_\mu\bar{\Psi})\gamma^\mu\Psi-\bar{\Psi}(\gamma^\mu\partial_\mu\Psi)\Big)-\bar{\Psi}(k-\gamma^\mu p_\mu)\Psi.
\end{equation*}
where $\bar{\Psi}=\zeta\Psi^\dagger$, with
\begin{equation*}
\zeta=-\frac{i}{\sqrt{2}}\{\gamma^4+\gamma^5\}=
\left(\begin{array}{cc}
0&-i\\
i&0
\end{array}\right),
\end{equation*}

In the case of Pauli-Schrödinger equation the association with the Wigner function is given by
\begin{equation*}
f_w= \Psi\star\bar{\Psi},
\end{equation*}
with each component satisfying Eq.~\eqref{W-L}.

\section{Gauge Theory for Non-Relativistic Spin $1/2$ particles in Phase Space}

The lagrangian density for non-relativistic spin $1/2$ particles in phase space can be written as
\begin{equation}\label{lag1}
\mathcal{L}=\overline{\Psi}\gamma^{\mu}\widetilde{(p_{\mu}\star)}\Psi - k\overline{\Psi}\Psi,
\end{equation}
where $A\widetilde{(p_{\mu}\star)}B=\frac{1}{2}[A(p_{\mu}\star B)-(p_{\mu}\star A)B]$. Our goal in this section is to analyze the invariance of Eq.(\ref{lag1}) under local gauge transformations given by
\begin{equation}\label{lag2}
\Psi=e^{-i\Omega}\star\Psi \qquad \overline{\Psi}=\overline{\Psi}\star e^{i\Omega},
\end{equation}
where $\Omega\equiv \Omega(q,p)$. For infinitesimal transformation, we have $\delta\Psi=-i\Omega\star\Psi$ and $\delta\overline{\Psi}=i\overline{\Psi}\star\Omega$, such that,
\begin{equation}\label{lag3}
\delta(p_{\mu}\star\Psi)=-ip_{\mu}\star\Omega\star\Psi,
\end{equation}
and 
\begin{equation}\label{lag4}
\delta(p_{\mu}\star\overline{\Psi})=ip_{\mu}\star\overline{\Psi}\star\Omega,
\end{equation}
It should be noted that $\delta(p_{\mu}\star\Psi)$ and $\delta(p_{\mu}\star\overline{\Psi})$ do not transform covariantly. For address this aspect we define the operator
\begin{equation}\label{lag5}
D_{\mu}\star=p_{\mu}\star - iA_{\mu}\star,
\end{equation}
leading to the modified lagrangian density
\begin{equation}\label{lag6}
\mathcal{L}=\overline{\Psi}\gamma^{\mu}\widetilde{(D_{\mu}\star)}\Psi - k\overline{\Psi}\Psi.
\end{equation}
Using the identity $p(f\star g)=f\star(pg)-\frac{i}{2}(\partial_{\mu}f)\star g$, the infinitesimal variation of $D_{\mu}\star\Psi$ is given by
\begin{equation}\label{lag7}
\delta(D_{\mu}\star\Psi)=-i\Omega\star(p_{\mu}\star\Psi)-\partial_{\mu}\Omega\star\Psi-A_{\mu}\star(\Omega\star\Psi)-i(\delta(A_{\mu})\star\Psi.
\end{equation}
Considering that $A_{\mu}$ transforms by
\begin{equation}\label{lag8}
A'_{\mu}\rightarrow A_{\mu}+i\{A_{\mu},\Omega\}_M + i \partial_{\mu}\Omega,
\end{equation}
where $\{a,b\}_M=a\star b - b \star a$ is the Moyal Brackets, we obtain
\begin{equation}\label{lag9}
\delta(D_{\mu}\star\Psi)=-i\Omega\star(D_{\mu}\star\Psi).
\end{equation}
Similarly, we have
\begin{equation}\label{lag10}
\delta(D_{\mu}\star\overline{\Psi})=-i(D_{\mu}\star\overline{\Psi})\star\Omega.
\end{equation}
In this sense, the Lagrangian density given in Eq.(\ref{lag6}) is invariant under transformation in Eq.(\ref{lag2}). Then, we have that the rule for minimal coupling is to replace $p_{\mu}\star$ by $D_{\mu}\star=p_{\mu}\star-iA_{\mu}\star$. In the next section we apply this formalism for analyze the Pauli-Schrödinger equation with electromagnetic interactions.

\section{Solution of the Pauli-Schrödinger Equation with Electromagnetic Interactions}\label{Spin 1/2 Symplectic Representaion}
The equation describing a spin $1/2$ particle in the Galilean covariant phase space is given by
\begin{eqnarray*}
\Bigg[\gamma^\mu\left(\widehat{P}_\mu-e\widehat{A}_\mu\right)-k\Bigg]\Psi.
\end{eqnarray*}
Making the following definition
\begin{eqnarray}
\Psi=\Big[\gamma^\nu\big(\widehat{P}_\nu-e\widehat{A}_\nu\big)+k\Big]\psi,
\end{eqnarray}
where $\widehat{P}_\nu=(p_\nu-\frac{i}{2}\partial_\nu)$.\vskip 2pt
Thus, we have
\begin{eqnarray}
\Big[\gamma^\mu\gamma^\nu\big(\widehat{P}_\mu-e\widehat{A}_\mu\big)\big(\widehat{P}_\nu-e\widehat{A}_\nu\big)-k^2\Big]\psi=0.\label{eq.g}
\end{eqnarray}
Considering $\gamma^\mu\gamma^\nu=g^{\mu\nu}+\sigma^{\mu\nu}$, where
\begin{eqnarray*}
\sigma^{\mu\nu}=\frac{1}{2}\Big(\gamma^{\mu}\gamma^\nu-\gamma^\nu\gamma^\mu\Big)=\frac{1}{2}[\gamma^\mu,\gamma^\nu].
\end{eqnarray*}
Using these results Eq.~\eqref{eq.g} becomes
\begin{eqnarray*}
\Bigg(\widehat{P}^\mu\widehat{P}_\mu-e\left(\widehat{P}^\mu\widehat{A}_\mu+\widehat{A}^\mu\widehat{P}_\mu\right)-e\sigma^{\mu\nu}\left[\widehat{P}_\nu,\widehat{A}_\mu\right]+e^2\widehat{A}^\mu\widehat{A}_\mu\Bigg)\psi=k^2\psi.
\end{eqnarray*}
Letting $\widehat{A}^i=\frac{1}{2}e^{ijk}B_j\widehat{Q}_k$, with $\widehat{Q}_\mu=(q_\mu+\frac{i}{2}\partial_{p^\mu})$ and $A^4=A^5=0$. Also, choosing the magnetic field as $\textbf{B}=(0,0,B)$.\vskip 2pt
Confining the motion of a particle to the plane $(q_1,q_2)$, i.e. $\widehat{P}_3=0$, we have the following equation
\begin{eqnarray}
&-&2\Bigg(p_4-\frac{i}{2}\partial_t\Bigg)\Bigg(p_5-\frac{i}{2}\partial_s\Bigg)\psi\nonumber\\
&+&\Bigg(p_1^2+p_2^2-\frac{1}{4}\Big(\frac{\partial^2}{\partial{q_1}^2}+\frac{\partial^2}{\partial {q_2}^2}\Big)-eB\Bigg[\frac{i}{2}\Big(p_2\partial_{p_1}-p_1\partial_{p_2}\Big)\nonumber\\
&+&\frac{1}{4}\Big(\frac{\partial^2}{\partial_{q_2}\partial_{p_1}}-\frac{\partial^2}{\partial_{q_1}\partial_{p_2}}\Big)\Bigg]-i\Big(p_2\partial_{q_2}+p_1\partial_{q_2}\Big)-eB\Bigg[(q_1p_2-q_2p_1)-\frac{i}{2}\Big(q_1\partial_{q_2}-q_2\partial_{q_1}\Big)\Bigg]\nonumber\\
&+&\frac{e^2B^2}{4}\Bigg[\Big(q_1+\frac{i}{2}\partial_{p_1}\Big)^2+\Big(q_2+\frac{i}{2}\partial_{p_2}\Big)^2\Bigg]-ie\sigma^{12}B\Bigg)\psi=k^2\psi,
\end{eqnarray}
where
$
\sigma^{12}=
i\left(\begin{array}{cc}
\sigma_3&0\\
0&\sigma_3
\end{array}\right).
$
\vskip 2pt
Letting
\begin{eqnarray*}
\psi=\left(\begin{array}{c}
\Phi(q^\mu,p^\mu)\\
\Theta(q^\mu,p^\mu)
\end{array}\right),
\end{eqnarray*}
we have two decoupled equations, one for $\Phi(q^\mu,p^\mu)$ and the other for $\Theta(q^\mu,p^\mu)$.
\begin{eqnarray*}
&-2\Bigg(p_4-\frac{i}{2}\partial_t\Bigg)\Bigg(p_5-\frac{i}{2}\partial_s\Bigg)\Phi(q^\mu,p^\mu)+\Bigg(p_1^2+p_2^2-\frac{1}{4}\Big(\frac{\partial^2}{\partial{q_1}^2}+\frac{\partial^2}{\partial {q_2}^2}\Big)-eB\Bigg[\frac{i}{2}\Big(p_2\partial_{p_1}-p_1\partial_{p_2}\Big)\\
&+\frac{1}{4}\Big(\frac{\partial^2}{\partial_{q_2}\partial_{p_1}}-\frac{\partial^2}{\partial_{q_1}\partial_{p_2}}\Big)\Bigg]-i\Big(p_2\partial_{q_2}+p_1\partial_{q_2}\Big)-eB\Bigg[(q_1p_2-q_2p_1)-\frac{i}{2}\Big(q_1\partial_{q_2}-q_2\partial_{q_1}\Big)\Bigg]\nonumber\\
&+\frac{e^2B^2}{4}\Bigg[\Big(q_1+\frac{i}{2}\partial_{p_1}\Big)^2+\Big(q_2+\frac{i}{2}\partial_{p_2}\Big)^2\Bigg]
+e\sigma^3B\Bigg)\Phi(q^\mu,p^\mu)=k^2\Phi(q^\mu,p^\mu),
\end{eqnarray*}
and equation for $\Theta$ is analogous.

Taking $\Phi(q^\mu,p^\mu)=\varphi(q^i,p^i)\phi(q^4,q^5,p^4,p^5)$. This give us the following equations
\begin{subequations}
\begin{eqnarray}
\Bigg(p_4-\frac{i}{2}\partial_t\Bigg)\Bigg(p_5-\frac{i}{2}\partial_s\Bigg)\phi=mE\phi+k^2\phi,\label{c_e1}
\end{eqnarray}
and
\begin{eqnarray}
&&\Bigg(p_1^2+p_2^2-\frac{1}{4}\Big(\frac{\partial^2}{\partial{q_1}^2}+\frac{\partial^2}{\partial {q_2}^2}\Big)-eB\Bigg[\frac{i}{2}\Big(p_2\partial_{p_1}-p_1\partial_{p_2}\Big)\nonumber\\
&+&\frac{1}{4}\Big(\frac{\partial^2}{\partial_{q_2}\partial_{p_1}}-\frac{\partial^2}{\partial_{q_1}\partial_{p_2}}\Big)\Bigg]-i\Big(p_2\partial_{q_2}+p_1\partial_{q_2}\Big)\nonumber\\
&-&eB\Bigg[(q_1p_2-q_2p_1)-\frac{i}{2}\Big(q_1\partial_{q_2}-q_2\partial_{q_1}\Big)\Bigg]\nonumber\\
\nonumber\\
&+&\frac{e^2B^2}{4}\Bigg[\Big(q_1+\frac{i}{2}\partial_{p_1}\Big)^2+\Big(q_2+\frac{i}{2}\partial_{p_2}\Big)^2\Bigg]+e\sigma^3B\Bigg)\varphi\nonumber\\
&=&2mE\varphi+k^2\varphi\label{c_e2}
\end{eqnarray}
\end{subequations}
The solution of eq.~\eqref{c_e1} is
\begin{eqnarray*}
\phi=C_1e^{-2i\left[(p_5+m)q_5+(p_4+E)t\right]},
\end{eqnarray*}
where $C_1$ is a normalization constant. To solve Eq.~\eqref{c_e2} we will make a changing of variables, defined by
\begin{eqnarray*}
w(q_1,q_2,p_1,p_2)&=&p^2_1+p^2_2+eB(q_2p_1-q_1p_2)\\
\\
&+&\frac{e^2B^2}{4}(q^2_1+q^2_2).
\end{eqnarray*}
After a long calculation, it is shown that the imaginary part of this equation is identically null, which gives us
\begin{eqnarray}
w\varphi-e^2B^2\frac{\partial\varphi(w)}{\partial \omega}-e^2B^2w\frac{\partial^2\varphi(w)}{\partial w^2}
=(2mE-esB+k^2)\varphi(w),
\end{eqnarray}
where $s\varphi=\sigma^3\varphi$, with $s=\pm 1$. Letting $\omega=w/(eB)$, $\alpha=(2mE-seB+k^2)/eB$ and defining $f(w)\equiv e^{w}\phi(\omega)$, we have
\begin{eqnarray}\label{eq_Kummer}
\omega f''(\omega)+(1-2\omega)f^\prime(\omega)-af(\omega)=0,
\end{eqnarray}

with $f^\prime(x)=\frac{\partial f}{\partial\omega}$ and $a=(1-\alpha)$. The equation~\eqref{eq_Kummer} is a confluent hypergeometric equation, more specifically the Kummer equation, and the physical solutions given by,
\begin{eqnarray*}
f_n(\omega)=A_nU\left(\frac{1}{2}-\frac{\alpha}{2},1,2\omega\right),
\end{eqnarray*}
where $U(a,b,x)$ are the Kummer's function and $A_n$ are constants. However, it is realized that if $a = -n$ with $n = 0,1,2, ..., $ the series $ U(a, b, x)$ becomes a polynomial series in $ x $ not exceeding n. Thus, writing,
\begin{eqnarray*}
\alpha-1=2n,
\end{eqnarray*}
we have the following relation of eigenvalue
\begin{eqnarray*}
E=\omega_c\left(n+\frac{1}{2}+\frac{s}{2}\right)-\frac{k^2}{2m},
\end{eqnarray*}
with $\omega_c=\frac{eB}{m}$ and corresponding the following auto-functions
\begin{eqnarray}
f_n(w)=A_nU\left(-n,1, \frac{2w}{eB}\right),
\end{eqnarray}
such that $ A_n $ are normalization constants. Therefore, the quasi-amplitudes become,
\begin{eqnarray}
\Phi_n=C_1e^{-2i\left[(p_5+m)q_5+(p_4+E)t\right]}
\Bigg(A_nU\left(-n,1, \frac{2w}{eB}\right)\exp\bigg(-\frac{w}{eB}\bigg)\Bigg).
\end{eqnarray}
The analog is valid for $\Theta$. These calculations provide the correct value of the eigenstates of energy for fermions in an external magnetic field, the well-known problem of Landau. Therefore, formalism is consistent with the standard calculation. To find the corresponding Wigner function just do $\psi_n\star\bar{\psi}_n$.\vskip 2pt
To find $ f_w $, we will do the same procedure for the harmonic oscillator. Just realize that $ w = 2mh $, with $$h=\frac{1}{2m}\Bigg(p^2_1+p^2_2+eB(q_2p_1-q_1p_2)+\frac{e^2B^2}{4}(q^2_1+q^2_2)\Bigg)$$.\vskip 2pt
Thus,
\begin{eqnarray*}
\psi_0=C_0 e^{-2h/\omega_c}
\end{eqnarray*}
Therefore
\begin{eqnarray*}
f_w^0=C_0 e^{-2h/\omega_c}\star\psi_0=C_0 e^{-2\hat{h}/\omega_c}\psi_0=C_0 e^{-2E_0/\omega_c}\psi_0
\end{eqnarray*}
Thus the ground state Wigner function for the spin particle $ 1/2 $ and $ -1 / 2 $ are given respectively
\begin{eqnarray*}
f_w^{0^+}&=&(C_{0^+})^2\frac{1}{e^2}e^{-(p^2_1+p^2_2+eB(q_2p_1-q_1p_2)+\frac{e^2B^2}{4}(q^2_1+q^2_2))/eB},\\
\nonumber\\
&\text{and}
\\
\\
f_w^{0^-}&=&(C_{0^-})^2e^{-(p^2_1+p^2_2+eB(q_2p_1-q_1p_2)+\frac{e^2B^2}{4}(q^2_1+q^2_2))/eB}.
\end{eqnarray*}
For the general case we have
\begin{eqnarray}
f_w^{n^\pm}=(A_n^{\pm})\Big(\frac{1}{n!\pi}\Big)e^{-(p^2_1+p^2_2+eB(q_2p_1-q_1p_2))} U\left(-n,1, \frac{2(p^2_1+p^2_2+eB(q_2p_1-q_1p_2))}{eB}\right)
\end{eqnarray}

The behavior of Wigner function for this system is shown in FIG1-Fig4. In graphics we notice that the Wigner function for the ground level, $n=0$, shown in FIG.1 does not present a negative part. However, others levels,  given in FIG.2-Fig4, present a negative part, and the graphics pointed out that the negative part of Wigner function is growing proportionally with levels of the system considered.
\subsection*{Non-classicality Indicator}
\vskip 8pt
The Wigner function satisfies the normalization condition $\int^{\infty}_{-\infty}f_w(q,p)dqdp=1$. Hence the doubled volume of the integrated negative part of the Wigner function can be written as
\begin{eqnarray*}
\eta(\psi)&=&\int^{\infty}_{-\infty}\Big(|f_w(q,p)|-f_w(q,p)\Big)dqdp\\
&=&\int^{\infty}_{-\infty}|f_w(q,p)|dqdp-1,
\end{eqnarray*}
so, $\eta(\psi)$ corresponds to a negativity indicator of the vector $|\psi\rangle$~\cite{kenfack2004}.\vskip 2pt
This indicator represents the doubled volume of the integrated part of the Wigner function. In sequence, we calculated numerically this indicator for the Landau levels. The results of this calculation are shown in Table~\eqref{neg}.
\begin{table}[!htbp]
\caption{Parameter of negativity of the Landau Levels for $n=1,2,3$.}
\label{neg}
\scalebox{1.4}{%
 \begin{tabular}{||c c c c c c c c c ||}
 \hline
$n$&\qquad&\qquad&\qquad\qquad&\qquad\qquad&\qquad\qquad&\qquad\qquad&\qquad&$\eta(\psi)$\\ [1ex]
 \hline
 0 &\qquad&\qquad&\qquad\qquad&\qquad\qquad&\qquad\qquad&\qquad\qquad&\qquad\qquad& 0 \\
 1&\qquad&\qquad&\qquad\qquad&\qquad\qquad&\qquad\qquad&\qquad\qquad&\qquad\qquad& 0.42612 \\
 2&\qquad&\qquad&\qquad\qquad&\qquad\qquad&\qquad\qquad&\qquad\qquad&\qquad\qquad& 0.72899 \\
3 &\qquad&\qquad&\qquad\qquad&\qquad\qquad&\qquad\qquad&\qquad\qquad&\qquad\qquad&0.97667 \\ [1ex]
 \hline
\end{tabular}
}
\end{table}
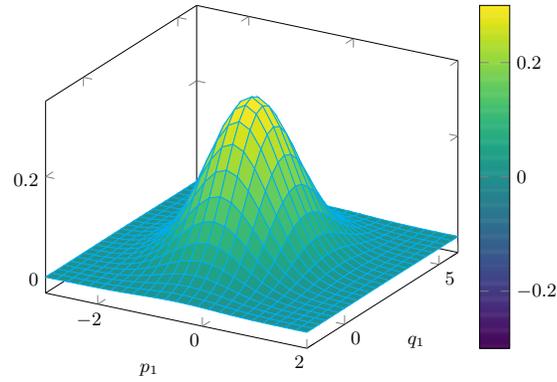
\begin{figure}[!htbp]
\centering
\begin{tikzpicture}[scale=0.8]
\begin{axis}[point meta min=-0.3, point meta max=0.3, colorbar, colormap/viridis,
    xlabel = $p_1$,
    ylabel = $q_1$,
    view={30}{30}
]
\addplot3[
    surf, faceted
color=cyan, domain=-3:2,y domain=-2:6
]
{1/pi*exp(-(x^2+1+(x-y)+1/4*(y^2+1))))};
\end{axis}
\end{tikzpicture}
\centering
\caption{Wigner Function (cut in $q_1$,$p_1$),Ground State.}
\end{figure}
\begin{figure}[!htbp]
\centering
\begin{tikzpicture}[scale=0.8]
\begin{axis}[point meta min=-0.3, point meta max=0.3, colorbar, colormap/viridis,
    xlabel = $p_1$,
    ylabel = $q_1$,
    view={30}{30}
]
\addplot3[
    surf, faceted
color=cyan, domain=-3:2,y domain=-2:6
]
{1/pi*((2*(x^2+1+(x-y)+1/4*(y^2+1)))-1)*exp(-(x^2+1+(x-y)+1/4*(y^2+1)))};
\end{axis}
\end{tikzpicture}
\centering
\caption{Wigner Function (cut in $q_1$,$p_1$),\\ First Excited State.}
\end{figure}
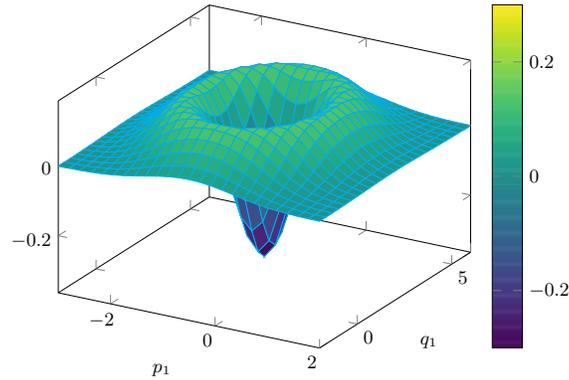
\begin{figure}[!htbp]
\centering
\begin{tikzpicture}[scale=0.8]
\begin{axis}[point meta min=-0.3, point meta max=0.3, colorbar, colormap/viridis,
    xlabel = $p_1$,
    ylabel = $q_1$,
    view={30}{30}
]
\addplot3[
    surf, faceted
color=cyan, domain=-3:2,y domain=-2:6
]
{1/(2*pi)*((4*(x^2+1+(x-y)+1/4*(y^2+1))^2-8*(x^2+1+(x-y)+1/4*(y^2+1))+2)*exp(-(x^2+1+(x-y)+1/4*(y^2+1)))};
\end{axis}
\end{tikzpicture}
\centering
\caption{Wigner Function (cut in $q_1$,$p_1$),\\ Second Excited State.}
\end{figure}
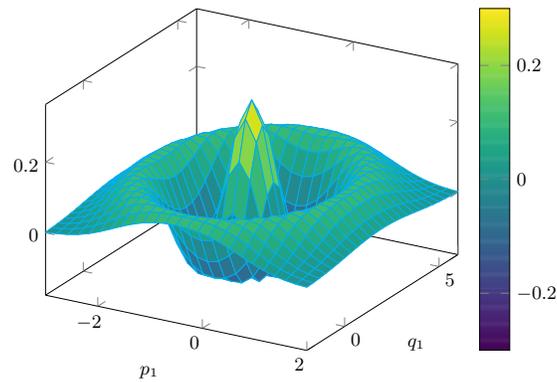
\begin{figure}[!htbp]
\centering
\begin{tikzpicture}[scale=0.8]
\begin{axis}[point meta min=-0.3, point meta max=0.3, colorbar, colormap/viridis,
    xlabel = $p_1$,
    ylabel = $q_1$,
    view={30}{30}
]
\addplot3[
    surf, faceted
color=cyan, domain=-3:2,y domain=-2:6
]
{1/(6*pi)*((8*(x^2+1+(x-y)+1/4*(y^2+1))^3-36*(x^2+1+(x-y)+1/4*(y^2+1))^2-36*(x^2+1+(x-y)+1/4*(y^2+1))-6)*exp(-(x^2+1+(x-y)+1/4*(y^2+1)))};
\end{axis}
\end{tikzpicture}
\centering
\caption{Wigner Function (cut in $q_1$,$p_1$),\\ Third Excited State.}
\end{figure}
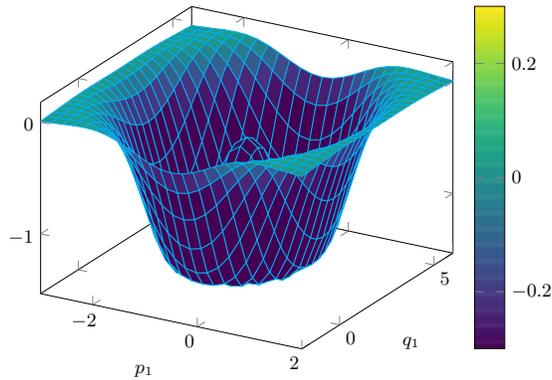
\pagebreak
\newpage

The results in Table~\eqref{neg} shown that the negativity indicator grows in the same direction
of the level $n$ of the system, in agreement with the graphics shown in FIG1-FIG4. This latter result can imply that the quantum entanglement
of quantum system analyzed increases when the indicator parameter grows. It is
important for quantum computing, for example see \cite{quantumcomputing}.

\section{Concluding Remarks}\label{Concluding Remarks}

We studied the spin $ 1/2 $ particle equation, the Pauli equation-Schrödinger, in the context of Galilean covariance and then construct a phase space formalism using such covariance. We begin with a presentation about the Galilean manifold where we use it to review the construction of Galilean covariance and the representations of quantum mechanics in this formalism, namely the spin $ 1/2 $ and scalar representation, equation of Schrödinger (Klein-Gordon-like) and the Pauli-Schrödinger (Dirac-like) equation respectively. \\
We construct the formalism of the quantum mechanics of phase space in the context of Galilean covariance and we arrive at the representations of the spin 0 and spin 1/2 equations, where for the spin equation 1/2, the Dirac-like equation. We analyzed the gauge symmetry for spin 1/2 particles in phase space and show that the minimal coupling in this case is obtained replacing in the lagrangian density $p_{\mu}\star$ by $p_{\mu}\star-iA_{\mu}\star$. We studied the electron in an external field and with the supposed solution we were able to recover the usual Pauli-Schrödinger equation (written non-covariant) in phase space. We also calculate the Wigner function for electron in an external field and implement the negativity parameter fir this system. As an important result, we have shown that the negativity indicator grows in the same direction
of the level $n$ of the system considered. 

\end{document}